\documentclass[twocolumn,showpacs,preprintnumbers,amsmath,amssymb]{revtex4}

\usepackage{graphicx}
\usepackage{dcolumn}
\usepackage{bm}

\begin{document}

\title{Scaling Mobility Patterns and Collective Movements: Deterministic Walks in Lattices}

\author{Xiao-Pu Han$^{1}$}
\author{Tao Zhou$^{1,2}$}
\author{Bing-Hong Wang$^{1,3}$}

\affiliation{$^{1}$Department of Modern Physics, University of Science and Technology of China, Hefei 230026, People's Republic of China\\ $^{2}$Web Sciences Center, University of Electronic Science and Technology of China, Chengdu 610051, People's Republic of China\\$^{3}$Research Center for Complex System Science, University of Shanghai for Science and Technology, Shanghai 200093, People's Republic of China}

\date{\today}

\begin{abstract}
Scaling mobility patterns have been widely observed for animals. In
this paper, we propose a deterministic walk model to understand the
scaling mobility patterns, where walkers take the least-action walks
on a lattice landscape and prey. Scaling laws in the displacement
distribution emerge when the amount of prey resource approaches the
critical point. Around the critical point, our model generates
ordered collective movements of walkers with a quasi-periodic
synchronization of walkers' directions. These results indicate that
the co-evolution of walkers' least-action behavior and the landscape
could be a potential origin of not only the individual scaling
mobility patterns, but also the flocks of animals. Our findings
provide a bridge to connect the individual scaling mobility patterns
and the ordered collective movements.



\end{abstract}


\maketitle

\section{Introduction}
Recently, the scaling properties in mobility patterns of animals
have attracted increasing attention \cite{RevV}. The traditional
scenario about the ``nearly random walks" of animals is now
challenged by the cumulated empirical observations, which indicate
heavy-tailed displacement distributions approximated to a power-law
form $P(l) \sim l^{-\alpha}$. Examples include the foraging process
and daily movements of wandering albatrosses \cite{visw1}, honeybees
\cite{reyn1, reyn2}, spider monkeys \cite{ramo}, microzooplanktons
\cite{bart1}, marine predators \cite{sims}, and so on. These
widespread observations imply a possible universal mechanism
underlying animals' mobility patterns.

Interpretations of the animals' mobility patterns can be divided
into two classes. One is the \emph{optimal search strategies}
\cite{visw2,bart2,reyn3,lom}, which indicate that animals can
maximize the searching efficiency by using power-law movements.
Another is the \emph{deterministic walks} (DW) \cite{boyer2, santo,
reyn5, boyer3} where a number of preys are randomly distributed on a
field, and a walker will continuously catch the nearest prey from
the current position. Recent studies introduced many real-life
factors into the standard framework of DW, such as olfactory-driven
foraging \cite{reyn4} and complex environment effects \cite{boyer1}.

To uncover the origin of scaling properties in animal mobility, we
propose a variant DW model that takes into account the regeneration
of resources in landscape and the least-action movements. Our model
can reproduce the power-law distribution of displacements, and the
scaling behavior in probability density of having moved a certain
distance at a certain time, agreeing well with the empirical
observations. In addition, our model generates ordered collective
movements of walkers with a quasi-periodic synchronization of
walkers' directions, indicating that the co-evolution of walkers'
least-action behavior and the landscape could be a potential
origin of not only the individual mobility patterns, but also the
population flocks.

\section{Model}

The food resources in the real environment can regenerate by
themselves as the growth and propagation of plants and preys, until
reaching a natural limitation of abundance. The maximization of
foraging benefits and minimization of costs (e.g., the least-action
movements) usually underlie the animals' behavior. Our model takes
into account these two fundamental ingredients. The environment is
represented by a two-dimensional $N \times N$ lattices with
non-periodical boundary condition (i.e., walkers can not go across
the boundary), and each lattice has prey resource $V(i,j)$ (for the
lattice at coordinate $(i,j)$). The maximum prey resource in each
lattice is set as a fixed value $V_m$. Different from the standard
DW, a more realistic case with multi-walkers (the number of walkers
is denoted by $M$) as well as their interactions are considered in
our model. The updating rules about the landscape and walkers'
positions are as follows:

(i) At each time step, each walker chooses the nearest lattice with
the maximum prey resource to occupy at the next time step, and if
there are more than one possible choices, the walker will randomly
select one (see Fig. 1). The movement is treated as instantaneous,
namely the diversity of velocity is ignored. The displacement (i.e.,
moving length) $l$ is defined as the geometric distance from the
current occupied lattice to the next occupied lattice, namely, $l =
\sqrt{(x_1 - x_0)^2+( y_1 - y_0)^2}$, where the coordinates
$(x_0,y_0)$ and $(x_1, y_1)$ respectively denote the current
position and the next position of the walker.

(ii) The resource $V$ in the current occupied lattice of the walker
is exhausted by the walker, namely $V\leftarrow 0$.

(iii) For each lattice with $V < V_m$ (currently not occupied by a
walker), the resource increases an unit at each time step until $V =
V_m$, representing the regeneration of prey resources.

(iv) When the number of walkers $M>1$, walkers update their
positions with random order asynchronously according to the above
algorithms at each time step.

\begin{figure}
\includegraphics[width=8.9cm]{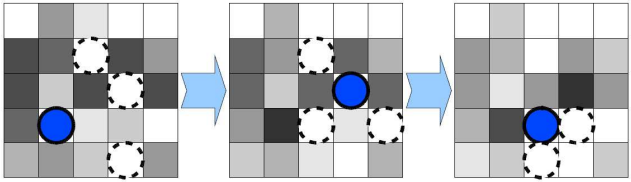}\\
\caption{(Color online) Illustration of the movement of a walker in
three successive time steps, where the blue fulled circle denotes
the current position of the walker, and white dashed circles denote
all the possible positions in the next time step. The values of $V$
in lattices are denoted by different gray scales, with pure-white
for $V=V_m$ and dark black for $V=0$.}
\end{figure}

Because the resource in each lattice regenerates with a fixed
speed, and each walker consumes at most $V_m$ resource at
each time step, we define $r = MV_m/S$ to express the ratio between
the total consumption of walkers and the total regeneration speed of
prey resource in the landscape, where $S = N \times N$ is the area
of the landscape. When $r = 1$, the consumption of prey resource is
equal to the regeneration speed, and the resource is in a critical
status. While if $r < 1$, the system has redundant resources. In our
simulations, $M$ and $r$ are tunable parameters, and the value
of $V_m$ is determined by $V_m = rS/M$.

Noticed that the movements in our model are not purely
``deterministic" when $r<1$ or $M>1$. This probabilistic property
comes from two sources: One is the random order in the updating of
walkers' positions when $M>1$, another is that a walker may have
more than one choices for the next position. Purely deterministic
case appears only when $r = 1$ and $M = 1$, in which the landscape
has only one lattice with  $V_m$ resource and the walker has to
periodically repeat its early trajectories. In the cases that
$r=1$ and $M>1$, although it is possible that walkers exchange
their trajectories under the treating with random order, the visited
time on each lattice is still strict periodic if $V_m$ is an
integer.

\section{Move length Distribution}

\begin{figure}
\includegraphics[width=8.7cm]{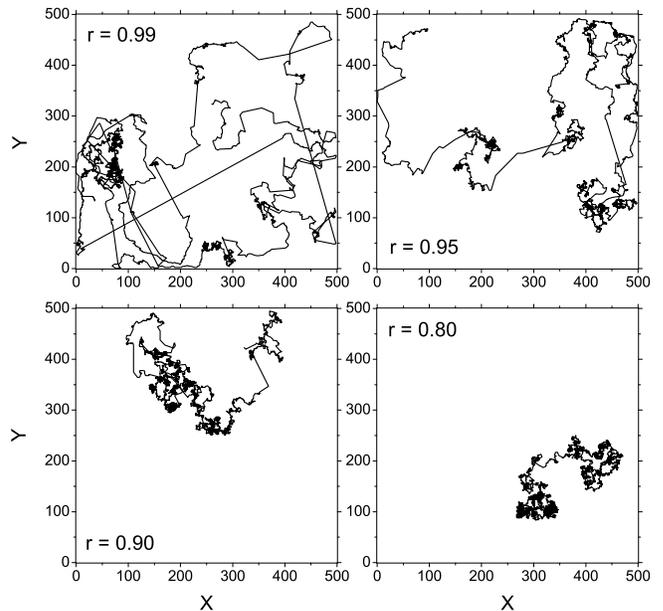}\\
\caption{The trajectories of a walker in 5000 consecutive steps for
different $r$. Other parameters are $M = 100$ and $N = 500$. }
\end{figure}

In our simulations, the size of landscape is fixed as $N=500$. We
assume the prey resource is full before a group of animals come
into their habitat, thus we set the initial prey resource of each
lattice to be $V_m$, and the initial positions of walkers are
randomly distributed in the landscape. Except for the case of
$r=1$, our simulations show that the move length distribution
gets stable after the evolution of $\frac{S}{(1.0-r)M}$ time
steps, so our statistics take into account the walkers' movement
after $\frac{S}{(1.0-r)M}$ time steps. When $r=1$, the number of
lattices having maximum resource is equal to the number of walkers
$M$ in the steady state. In this case, each walker has to repeat
its early trajectory or other walkers' after the first $S/M$ time
steps (after that time, the consumption equals regeneration), so our statistics take into account the walkers' movement
after $S/M$ time steps.

Figure 2 shows the trajectories of a walker for different $r$,
where abundant long-range movements can be observed when $r$
approaches to $1$. The move length distributions $P(l)$ for
different $r$ and $M$ are respectively shown in Fig. 3(a) and
3(b). When $r$ approaches 1, a scaling property of the move
length distribution, say $P(l)\sim l^{-\alpha}$, can be observed.
The analytical result for the move length distribution is given in the Appendix, which agrees with the simulations.
As mentioned above, when $r=1$, the trajectory is periodic and thus at the last time step of a period, the walker returns to its origin, corresponding to a generally long displacement with the same order of the system size. Therefore, as shown in Fig. 3(a), when $r=1$, a peak appears at $l \approx N = 500$. The dependence between the power-law exponent $\alpha$ and the
parameters $r$ and $M$ are shown respectively in the two inserts
of Fig. 3(a) and 3(b). Except $r=1$, $\alpha$ decreases
monotonously with the increasing of $r$. For example, when
$M=100$, $\alpha$ decreases from $3.3$ to $2.2$ when $r$ changes
from $0.80$ to $0.99$. This range of $\alpha$ covers almost all
the known real-world observations of move length distributions of
animals. This result indicates that the walker is more likely to
take long-range movement when the prey resource is not rich
enough, which is in accordance with the experiment on the prey
behavior of bumble-bees \cite{visw2} and also is supported by the resent observation on the movements patterns of marine predators \cite{Hum}. Our result suggests that the
animals living in a habitat with abundant prey resource will not
display scaling property in their mobility.

\begin{figure}
\includegraphics[width=8.7cm]{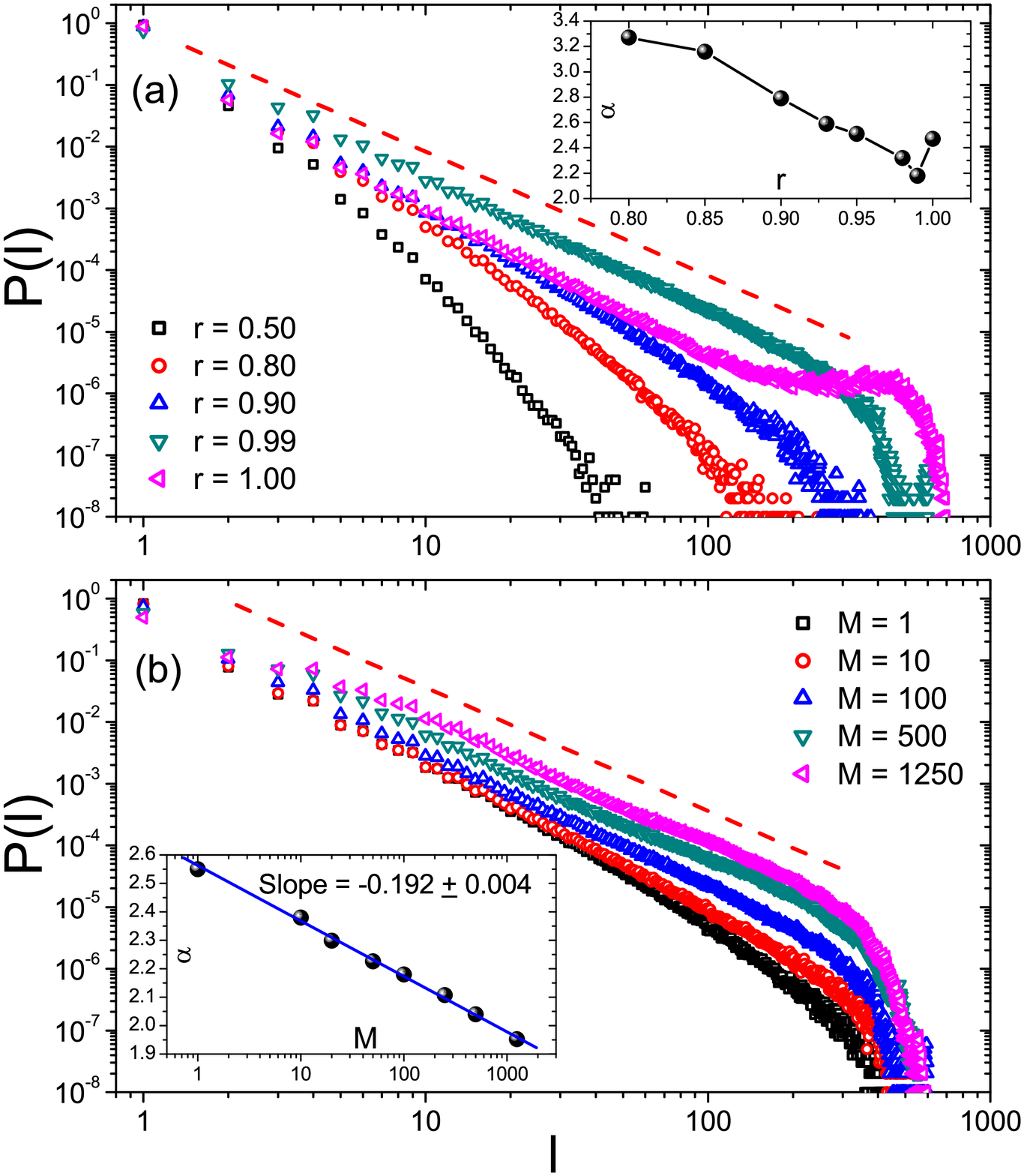}\\
\caption{(Color online) (a) Move length distribution $P(l)$ for
different $r$, where the inset shows the dependence between the
power-law exponent $\alpha$ and $r$. Other parameters are $M=100$
and $N=500$. (b) $P(l)$ for different $M$, where the inset shows
the dependence between $\alpha$ and $M$, and the blue line denotes
the fitting line with slope $-0.192 \pm 0.004$. Other parameters
are $r=0.99$, and $N=500$. The red dashed lines in the two plots
represent a power law with exponent $-2$. All the data points are
averaged by 100 independent runs, each of which includes $10^6$
movements. The size of error bars in the two insets are smaller
than the data points.}
\end{figure}


As shown in the inset of Fig. 3(b), the relation between $\alpha$ and $M$ can be well captured by logarithmic form $\alpha \sim -\ln M$. The case of  $M = 1$ corresponds to a solitary animal living in a fixed territory, while $M>1$ represents the case where several animals share prey resource in the same field. This result indicates that the individuals in large group are more likely to take long-range movements when the resource is not sufficiently rich.




\section{Collective Movements}

\begin{figure}
\includegraphics[width=6.0cm]{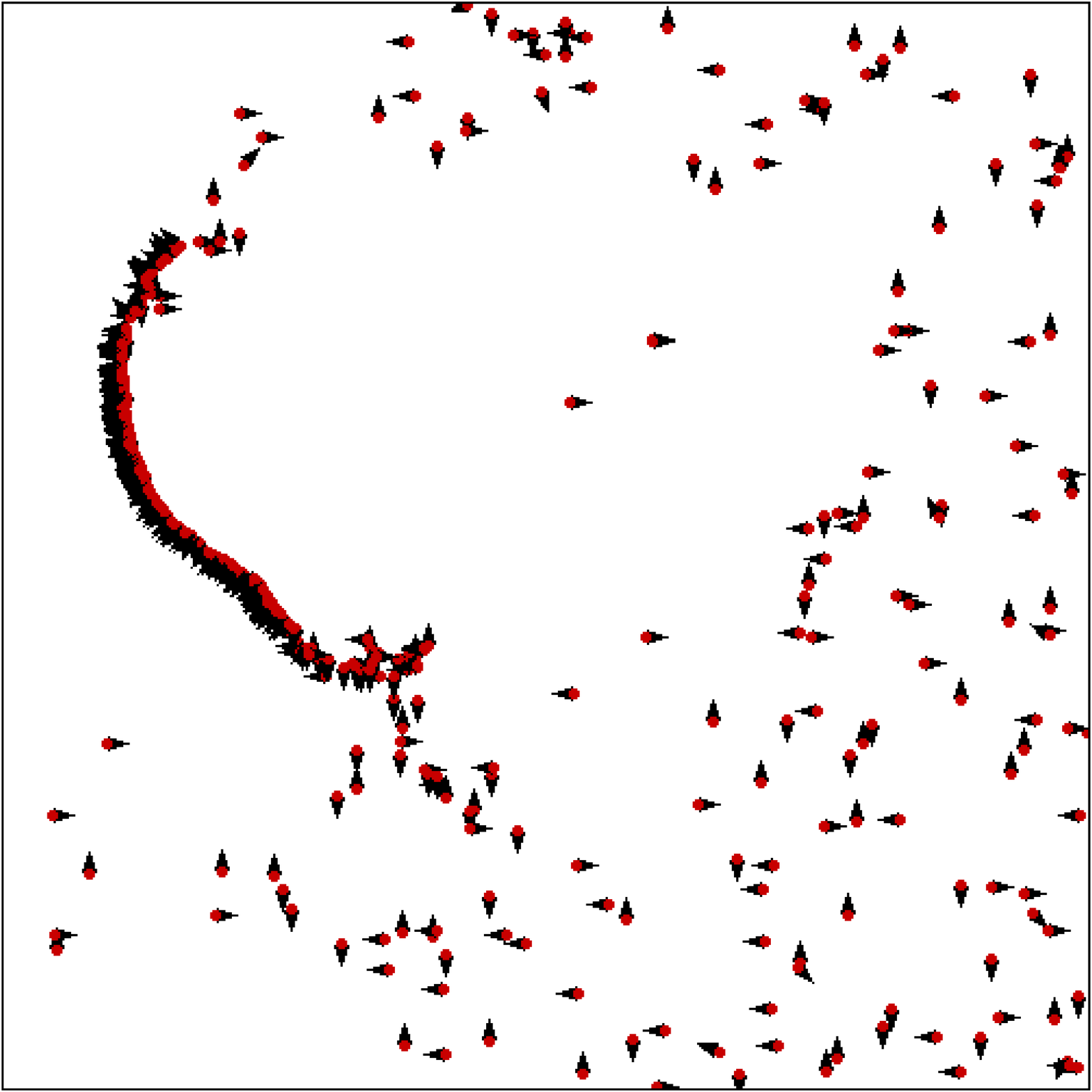}\\
\caption{(Color online) A typical ordered marching band of walkers generated by our model when $t = 50425$ ($\psi = 0.57$). The red circles and black arrows denote the current positions and directions of each walkers respectively. The parameters are $r = 0.99$, $M = 500$ and $N = 500$. }
\end{figure}

\begin{figure}
\includegraphics[width=8.9cm]{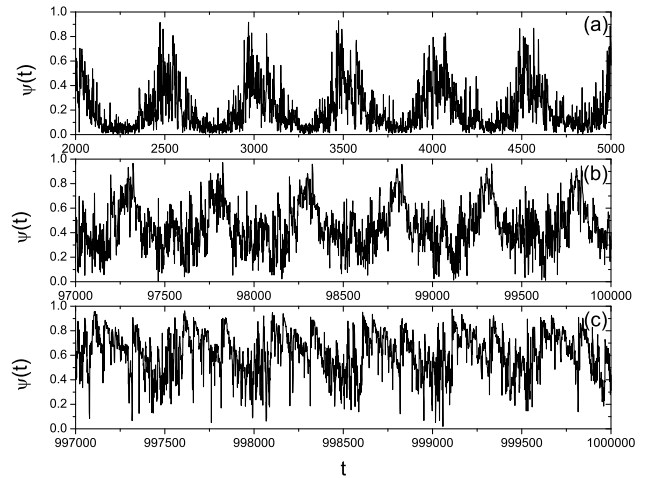}\\
\caption{The periodic varying of order parameter $\psi(t)$ in three time periods with parameters $r = 0.99$, $M = 500$, and $N = 500$. }
\end{figure}

To our surprise, collective movements are observed when $M$ is large and $r$ is close to 1, where walkers may line up one or several marching bands, and the walkers in the same marching band have similar moving directions (approximately perpendicular to the band). A typical example is shown in Fig. 4. Sometimes over half walkers are in these marching bands. Marching bands are not stable. They may suddenly emerge or disappear, may grow larger or fall into pieces.



We define an order parameter $\psi(t)$ to measure the degree of synchronization of walkers' directions at time step $t$ as the following form \cite{Vics}:
\begin{equation}
\psi(t) = \frac{|\sum{\vec{v}_i(t)}|}{\sum{|\vec{v}_i(t)|}},
\end{equation}
where the velocity vector $\vec{v}_i(t) = (x_{t+1}-x_t)\hat{x}+(y_{t+1}- y_t)\hat{y}$, and coordinate $(x,y)$ denotes the position of the $i$-th walker, $\hat{x}$ and $\hat{y}$ respectively denote the unit of velocity on the direction $x$-axis and $y$-axis. Obviously, $\psi = 1$ if all walkers have the same direction.

Figure 5 reports three typical examples of the $\psi(t)$ curves, respectively for early, middle and relatively later stages, where the quasi-periodic behavior with period length about $S/M$ can be observed. As shown in Fig. 6, when $r = 0.99$, the order parameter will exceed $0.6$, indicating a strong synchronization of individual's directions. Collective movements can be only observed when $r$ is close to $1$, which is also the condition for the emergence of scaling mobility patterns as mentioned in Section III. This property is demonstrated by the simulations shown in Fig. 6, where one can see that the steady value of order parameter is very sensitive to the parameter $r$, and when $r$ goes down from $1$, the collective movements sharply disappear.

Notice that our model does not imply any direct interaction between walkers. They are driven by the co-evolution of resource landscape and the least-action movements. This feature is far different from most known interpretations on the dynamical mechanisms of animal collective movements \cite{Vics,Couz1,Couz2,Nagy,Ball}, yet similar to the so-called active walk process \cite{Lam1,Lam2}, where the macroscopic-level structure emerges from the interplay between walkers and landscape. The existence of synchronized motions is very sensitive to the value of $r$, suggesting that the food shortage may be responsible to the emergence of animal collective behaviors, which has also been observed for locusts \cite{Buhl}. The existence of the scaling law in displacement distribution and the collective movements are, to our surprise, under the same condition $r \rightarrow 1$. However, these two phenomena are not the two sides of a coin, actually, they do not straightforwardly depend on each other. Whether there exists a certain mechanism underlying the co-existence is still an open question to us.



\begin{figure}
\includegraphics[width=8.9cm]{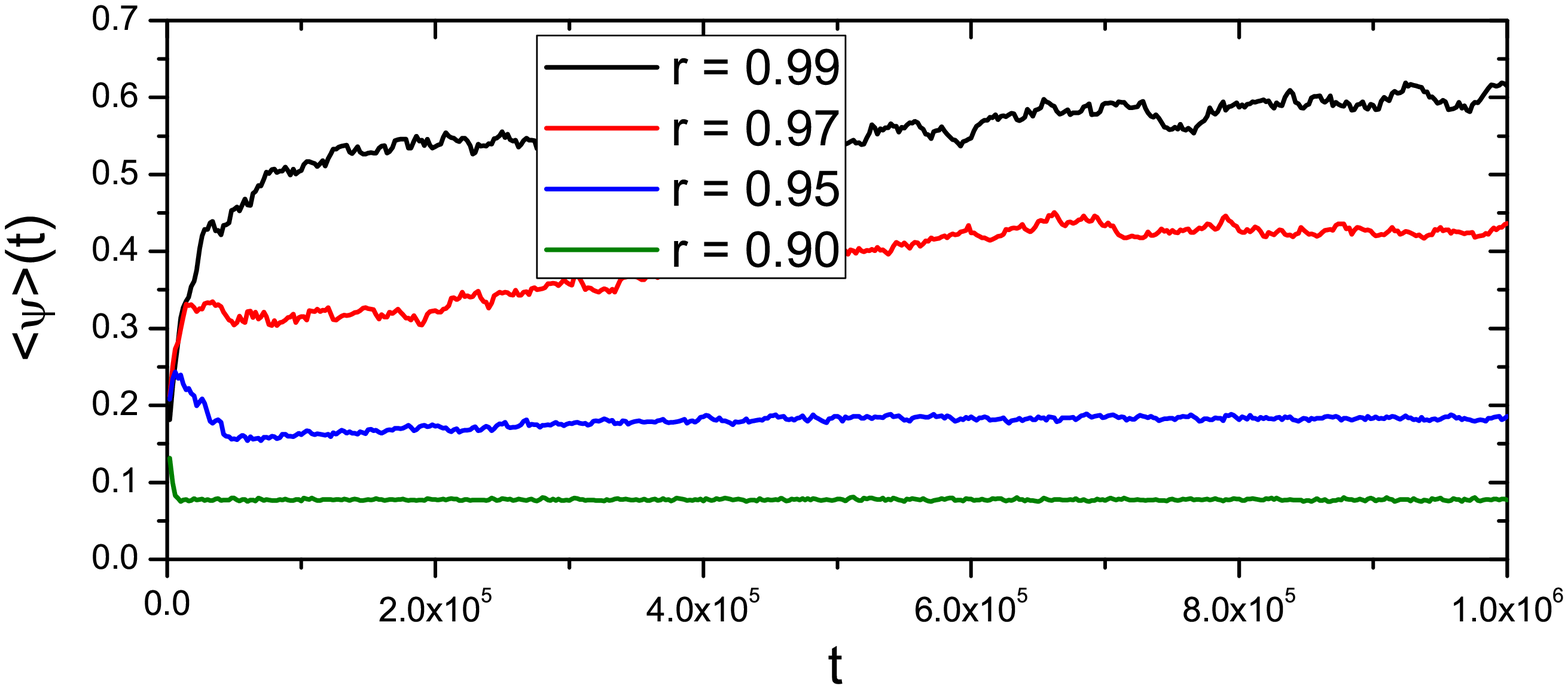}\\
\caption{(Color online) The average value in each period of order parameter $\left\langle \psi\right\rangle (t)$ for different $r$, namely $\left\langle \psi\right\rangle (t)$ is the average value of $\psi$ in the range $[t-\frac{S}{2M}, t+\frac{S}{2M}]$. Simulations run with parameter setting $M = 500$ and $N = 500$. All the data points are averaged by 10 independent runs.
}
\end{figure}

\begin{figure}
\includegraphics[width=8.9cm]{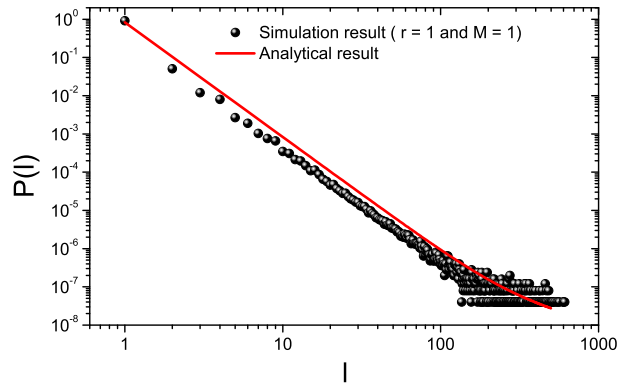}\\
\caption{(Color online) Comparison of the analytical result (Eq.
(A5), red line) and the simulation result (black circles) when $r =
1$, $M = 1$, and $N = 500$. The term $l^{-1}$ plays a role only for
large $l$.}
\end{figure}

\section{conclusions and discussions }

Our model mimics the mobility patterns of many least-action walkers
which prey in an landscape with regenerating ability. Scaling laws
of move length distribution emerges when the regeneration speed of
prey resource approaches to the critical point that the amount of
resource is just enough. This result indicates that the mobility
patterns of animals are sensitive to the environmental context (e.g.
food resource), which is qualitatively supported by real
observations \cite{visw2,Hum}. Our model indicates that population
also highly affects on the mobility patterns (see the inset of Fig.
3(b)), which is rarely discussed in early DW models. In addition,
our model generates quasi-periodic collective movements with
marching bands when $r \rightarrow1$ indicating that the
aggregations of animals are more likely to appear with food shortage
\cite{Buhl,Hend}, which is far different from many known dynamical
interpretations based on the interaction between individuals
\cite{Vics,Couz1,Couz2,Nagy,Ball}. One of the noticeable features in
our results is the coexistence of both scaling mobility patterns in
individual level and collective movement in population level. Both
phenomena are under almost the same condition: the amount of prey
resource approaches to the critical point, implying that the
environment-driven mechanism may bridge the scaling individual
activity patterns and global ordered behaviors.


Under different parameter settings, our results exhibit a wide and gradual spectrum of different mobility features: from the scaling movements to the random-walk-like mobility pattern, from ordered collective movements to the uncorrelated motions. These results are respectively well in agreement with different types of real-world mobility patterns of animals.
However, not all the results in our model are fully addressed. Some phenomena, such as the logarithmic relation between $\alpha$ and $M$, and the microscopic mechanism in the emergence of such collective movements, are still open questions to us.


In a word, although our model is a minimum model that many real factors such as the diversity of velocity rates and the heterogeneity of environment are not considered, the results of our model are generally in agreement with many wild observations. Our model could be helpful in the understanding of the origin of both scaling mobility patterns and the ordered collective movements of animals.

\begin{acknowledgments}
This work was supported by the National Natural Science Foundation
of China Grants Nos. 70871082, 10975126, 70971089 and 10635040.
\end{acknowledgments}

\appendix

\section{Mean-field Analysis}

When $r\rightarrow 1$, the analysis can be simplified by the
periodic movements. At the critical point, the period length is
equal to $S/M$ and during a period, each lattice will be visited
exactly once. We introduce a mean-field approximation that at any
time, the unvisited lattices are evenly distributed in the space.
After $\tau$ time steps, the number of unvisited lattices is
$m(\tau)=S-M\tau$. Defining the normalized move length $l_* = l/N =
lS^{-1/2}$ where $l$ is the real geometric length, the normalized
probability density of the distance (move length) from an unvisited
lattice to its nearest unvisited lattice after $\tau$ steps is:
\begin{equation}
p(l_*,\tau) \approx [m(\tau)-1] \times 2\pi l_* (1 - \pi
l_*^2)^{m(\tau)-2}.
\end{equation}
The move length distribution $P(l_*)$ during a period is the
cumulation of these $p(d,\tau)$:
\begin{equation}
P(l_*) \approx \frac{M}{S}\int_{0}^{S/M}p(l_*,\tau)d\tau.
\end{equation}
Substitute $m(\tau)=S-M\tau$ into Eq. (A1) and Eq. (A2),  $P(l_*)$ can be obtained as:
\begin{equation}
P(l_*) \approx \frac{2\pi l_*}{S}[{(\ln a)}^{-2}+{(\ln
a)}^{-1}]a^{-2},
\end{equation}
where $a = 1-\pi l_*^2$. Mostly $l_*\ll 1$ and thus $\ln(1-\pi
l_*^2) \approx \pi l_*^2$ and $a^{-2}\approx 1$. Therefore $P(l_*)$
can be written as
\begin{equation}
P(l_*) \approx \frac{2}{S}(\frac{1}{\pi} l_*^{-3}+l_*^{-1}),
\end{equation}
corresponding to the distribution
\begin{equation}
P(l) \sim (\frac{S}{\pi} l^{-3}+l^{-1}).
\end{equation}

The range of $l$ is limited from $1$ to $\sqrt{2S}$, therefore if
$S$ is very large, the distribution $P(l)$ is mainly determined by
the first term $\frac{S}{\pi}l^{-3}$. Figure 7 reports the
analytical result Eq. (A5), which agrees with the simulation well.
Notice that, the term $l^{-1}$ only show its effect for very large
$l$.

\end{document}